\documentclass[twocolumn,english,fp]{jpsj3}

\usepackage{units}
\usepackage{ulem}
\usepackage{upgreek}
\usepackage{babel}

\begin{document}
\title{Ferromagnetic Quantum Criticality Studied by Hall Effect Measurements in UCoAl}

\inst{\address{
$^{1}$INAC, SPSMS, CEA Grenoble, 38054 Grenoble, France}\\
$^{2}$Institute for Materials Research, Tohoku University, Oarai, Ibaraki, 311-1313, Japan
}

\abst{Hall effect measurements were performed under pressure and magnetic field up to $\unit[2.2]{GPa}$ and $\unit[16]{T}$ on a single crystal of UCoAl. At ambient pressure, the system undergoes a first order metamagnetic transition at the critical field $B_{m}=\unit[0.7]{T}$ from a paramagnetic ground state to a field--induced ferromagnetic state. The Hall signal is linear at low field and shows a step--like anomaly at the transition, with only little change of the Hall coefficient. The anomaly is sharpest at the temperature of the critical end point $T_{0}=\unit[12]{K}$ above which the first order metamagnetic transition becomes a crossover. Under pressure $B_{m}$ increases and $T_{0}$ decreases. The step--like anomaly in the Hall effect disappears at $P_M \approx \unit[1.3]{GPa}$ and the metamagnetic transition is not detected above the quantum critical end point (QCEP) at $P_{\Delta}\approx\unit[1.7]{GPa}$, $B_{m}\approx\unit[7]{T}$. Using magnetization data, we analyse our Hall resistivity data at ambient pressure in order to quantitatively account for both ordinary and anomalous contributions to the Hall effect. Under pressure, a drastic change in the field dependence of the Hall coefficient is found on crossing the QCEP. A possible Fermi surface change at $B_m$ remains an open question.}

\kword{quantum critical end point, anomalous Hall
effect, metamagnetism, ferromagnetism, UCoAl}

\author{Tristan COMBIER$^{1}$\thanks{Email: Tristan.Combier@cea.fr}, Dai AOKI$^{1,2}$\thanks{Email: Dai.Aoki@cea.fr}, Georg KNEBEL$^{1}$, and Jacques FLOUQUET$^{1}$}

\maketitle

\section{Introduction}

The metamagnetism in strongly correlated electron systems with Ising--type ferromagnetic (FM) behaviour is intensively studied because it produces a variety of unconventional effects. In some itinerant ferromagnets, such as $\mathrm{UGe_{2}}$ \cite{Taufour2010,Kotegawa2011} or $\mathrm{ZrZn_{2}}$ \cite{Uhlarz2004}, one can drive the Curie temperature $T_C$ to $\unit[0]{K}$ by tuning an external control parameter like pressure, and the ground state is found to be paramagnetic (PM) above the quantum critical point (QCP). In theory, it has been suggested ---~and in some cases experimentally shown~--- that the second order ferromagnetic transition changes to first order at a tricritical point\cite{Belitz2005}. By applying a magnetic field above the critical point in the paramagnetic phase, such systems eventually recover their ferromagnetic state by undergoing a first--order metamagnetic transition at $B_m$, drawing a wing--shape first order transition plane in the temperature ($T$) -- pressure ($P$) -- field ($B$) phase diagram. The first order transition terminates at high pressure and high field at $T=0$ at the so--called quantum critical end point (QCEP). In this critical region, only few experiments were carried out because of the severe experimental conditions of low temperature, high pressure, and high field \cite{Taufour2010,Kotegawa2011,Uhlarz2004,Wu2011}. Due to the recent focus on quantum criticality, the metamagnetism in itinerant ferromagnets has been recently revisited theoretically (see e.g.~Ref.~\citen{Zacharias2013,Bercx2012}). The main debate is whether a Lifshitz--like transition is associated with the occurence of metamagnetism.

A good candidate to investigate itinerant metamagnetism is the heavy fermion compound UCoAl, with ZrNiAl--type hexagonal structure, space group P$\bar{6}$2m. At ambient pressure, its ground state is paramagnetic, with strong uniaxial magnetic anisotropy. By applying magnetic field along the easy magnetization axis (\textit{c}--axis), a sharp first--order metamagnetic transition occurs at low temperature at a critical field $B_{m}\sim\unit[0.7]{T}$ \cite{Andreev1985,Mushnikov1999}. The first--order nature of the metamagnetic transition terminates at a critical end point (CEP) at $T_0\sim\unit[12]{K}$, and changes into a crossover at higher temperature \cite{Aoki2011, Karube2012, Palacio-Morales2013}.
By applying pressure, recent resistivity and magnetostriction experiments\cite{Aoki2011} showed that the critical field increases up to the QCEP, located at $P_{QCEP}\sim\unit[1.6]{GPa}$ and $B_{QCEP}\sim\unit[7]{T}$, where an acute enhancement of the effective mass $m^\star$ of the quasiparticles has been detected.
Previous magnetization experiments under pressure are in good agreement with an initial pressure increase of  $B_m$\cite{Mushnikov1999}. It is worthwhile to remark that just above $P_{QCEP}$  in the paramagnetic ground state in high magnetic field, sharp pseudometamagnetism at $B_m$ will replace the sharp first order metamagnetism below $P_{QCEP}$\cite{Zacharias2013}. Further increasing pressure must lead to a broadening of the pseudometamagnetic transition.

An interesting tool to unveil the physical properties in the vicinity of a QCEP is the Hall effect as it has two main contributions: the normal Hall effect (NHE), which is linked to the type of carriers and their number, and the anomalous Hall effect (AHE), which is linked to the magnetization $M$. Changes in the effective mass ($m^*$) associated (or not) to a Fermi surface instability at the QCEP may have a specific signature.

In this article we present detailed Hall effect measurements performed on a single crystal
of UCoAl, at low temperatures down to $\unit[150]{mK}$, under hydrostatic pressure up to $\unit[2.2]{GPa}$ and magnetic field up to $\unit[16]{T}$.

\section{Experimental}

Single crystals of UCoAl were prepared from depleted U (99.9\% -- 3N), Co (3N) and Al (5N) in stoichiometric proportions. The components were melt in a tetra--arc furnace under Ar gas atmosphere and a single crystal ingot was pulled by Czochralski method with pulling rate of $\unit[15]{mm/h}$. The single crystal was checked by X--ray Laue photograph. A flat $c$--plane rectangular sample of $1.8\times1.2\times\unit[0.2]{mm^{3}}$ was cut with a spark--cutter for the Hall effect measurements. Its residual resistivity ratio $RRR=14$ is within the highest reported for UCoAl indicating the very good quality of the crystal.

Magnetization measurements at ambient pressure were performed on another sample from the same ingot, with a commercial SQUID magnetometer (Quantum Design) down to $\unit[1.8]{K}$ and up to $\unit[5.5]{T}$. The metamagnetic critical field $B_m = 0.6$~T of that crystal is slightly lower than that of the crystal used for the Hall effect experiment where we found $B_m = 0.7$~T. This shows the high sensitivity of $B_m$ to the quality and/or homogeneity of the single crystals.

The Hall effect $\rho_{xy}$ was measured using a four--probe AC lock--in technique. The current was applied perpendicular to the \textit{c}--axis in the hexagonal planes (typically $\unit[1]{mA}$ at a frequency $f\sim\unit[17]{Hz}$) and positive and negative fields were applied along the \textit{c}--axis, which is the easy magnetization axis, in order to cancel out contributions of the magnetoresistance  $R_{xx}$ due to a small misalignement of the transverse voltage contacts. The symmetric magnetoresistance $R_{xx}$ has been analysed too and good agreement with previous resistivity measurements \cite{Aoki2011} has been observed. The sample was set in a hybrid CuBe -- NiCrAl piston cylinder cell for the pressure studies with Daphne oil 7373 as pressure transmitting medium. The pressure was determined from the superconducting transition temperature of Pb by AC susceptibility measurements. The Hall effect measurements were performed both in a PPMS (Quantum Design) ($2$ to $\unit[300]{K}$ ; $0$ to $\unit[9]{T}$) and in a homemade dilution refrigerator ($0.15$ to $\unit[10]{K}$ ; $0$ to $\unit[16]{T}$).

\section{Hall effect at ambient Pressure}

\begin{figure} 
	\centering
		\includegraphics{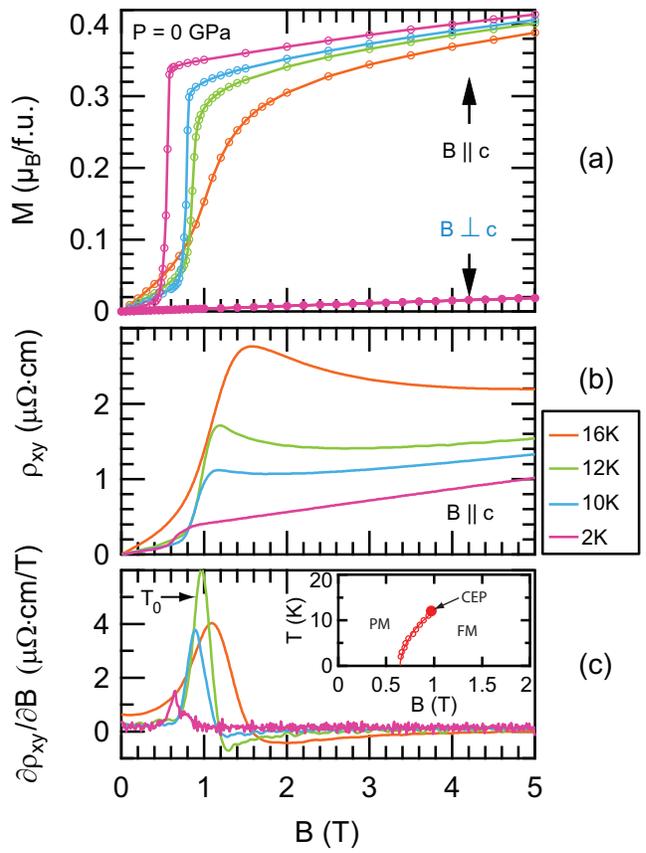}
		\caption{\label{ambient_P} \textit{(color online)} (a) Magnetization ($B\parallel$ and $\perp c$--axis), (b) Hall effect $\rho_{xy}$ ($B \parallel c$), and (c) $\partial\rho_{xy}/\partial B$ vs $B$ at ambient pressure at various temperatures ($2, 10, 12, \unit[16]{K}$). The temperature of the critical end point of the first order transition $T_{0}=\unit[12]{K}$ is taken as the maximum of the derivative in (c). The inset in (c) shows the ($T,B$) phase diagram at ambient pressure from the Hall effect.}
\end{figure}

Figure \ref{ambient_P}(a) shows the field dependence of the magnetization at ambient pressure for $B\parallel c$ and $B\perp c$--axis at various temperatures. At low temperatures ($T=2$~K) a sharp metamagnetic transition is observed for $B\parallel c$ at $B_{m}\approx\unit[0.7]{T}$ from a paramagnetic ground state to a polarized ferromagnetic state with an ordered moment $M_{0}\approx\unit[0.3]{\mu_{B}/U}$. The magnetization for $B\perp c$ increases linearly and  $M \approx\unit[0.023]{\mu_{B}/U}$ is found for $B = \unit[5]{T}$, indicating the strong Ising--type anisotropy.

The metamagnetic transition at $B_m$ is of first order, attested by a hysteresis between increasing and decreasing field magnetization curves \cite{Andreev1985, Aoki2011}. The hysteresis reduces and the metamagnetic transition gradually broadens as temperature increases, until it ends at $T_{0}=\unit[12]{K}$, marking the first--order critical end point (CEP). For $T > T_0$ a pseudo--metamagnetic transition fades out into a broad crossover region \cite{Palacio-Morales2013}.

The Hall resistivity $\rho_{xy}(B)$, see Fig.~\ref{ambient_P}(b), is positive through the entire magnetic field range  along the $c$ axis. In difference, a negative Hall resistivity has been reported for $B \perp c$ \cite{Matsuda2000} indicating the multi--band character.  At the lowest temperatures, the metamagnetic transition appears as a step--like increase in  $\rho_{xy}(B)$. This feature is a clear illustration of the interplay between $\rho_{xy}$ and the magnetization. With increasing temperatures ($\unit[4]{K}<T<T_0\sim\unit[12]{K}$), a peak develops just above the critical field $B_m$. Further increasing temperature, the step--like increase at $B_m$ fades out and only a broad maximum is observed in $\rho_{xy} (B)$ above $T_0$.
Similar features were also found in transverse magnetoresistance $\rho_{xx}(B)$ \cite{Aoki2011}. This peak in $\rho_{xx}$ was interpreted as an enhancement of the magnetic scattering contribution due to strong magnetic fluctuations around the CEP \cite{Nohara2011}.

From the field derivative of $\rho_{xy}$, the differential Hall constant $\tilde{R}_{H}=\partial\rho_{xy}(B)/\partial B$, see Fig.~\ref{ambient_P}(c), we locate the critical field $B_m$  at the peak in $\partial\rho_{xy}(B)/\partial B$. The height of the peak is maximal at $T_0=\unit[12]{K}$, while the width does not broaden with increasing temperature until $T_0$. Above $T_0$, the peak height decreases and the width is strongly increasing. We will use the same criterion to determine the first--order critical end point at $(B_m^\star,T_0)$ under pressure. Let us note that the differential Hall coefficient $\tilde{R}_H$ shows little change across $B_m$ (15\% decrease), going from $\unit[0.18]{\upmu\Omega\cdot cm/T}$ in the PM region to $\unit[0.15]{\upmu\Omega\cdot cm/T}$ in the FM region.
By plotting $\partial\rho_{xy}(H,T)/\partial B$ we can draw the $(T,H)$ phase diagram of UCoAl at ambient pressure (see inset in Fig.~\ref{ambient_P}(c)) in excellent agreement with previous work \cite{Aoki2011}. The crossover region extends at least up to $\unit[20]{K}$ in the continuity of the metamagnetic transition.

\begin{figure} 
	\centering
		\includegraphics{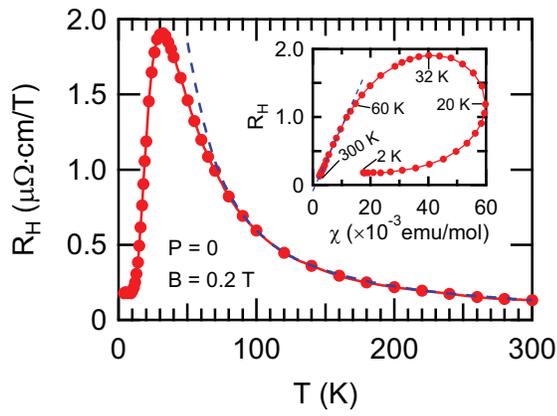}
	
	\caption{\label{Rh_noP} \textit{(color online)} Temperature dependence of the Hall coefficient $R_H$ for $B=\unit[0.2]{T}\parallel c$--axis (full circles). The dotted line is a high temperature fit using magnetic susceptibility data (dotted line). The inset shows $R_H$ plotted against the magnetic susceptibility. The linear dependence indicates that the high temperature regime is dominated by the AHE.  We obtain $R_0=\unit[-0.1]{\upmu\Omega\cdot cm/T}$ and $R_S=\unit[97]{\upmu\Omega\cdot cm/T}$.}
\end{figure}

Figure \ref{Rh_noP} shows the temperature dependence of the linear Hall coefficient $R_H=\rho_{xy}/B$ at low field ($B=\unit[0.2]{T}$). 
As we can see, $R_H$ has a maximum around $\unit[30]{K}$, connected with a maximum of the magnetic susceptibility $\chi$ at $T=\unit[20]{K}$. The inset of Fig.~\ref{Rh_noP} shows $R_H$ plotted against $\chi$ with the temperature as an incipient parameter. $R_H$ is proportional to the magnetic susceptibility $\chi$ from 300~K down to 70~K. In general, the Hall effect can be expressed as the sum of NHE proportional to the carrier contribution  and an anomalous  contribution from left--right asymmetric scattering due to the ordered magnetic moments (AHE):
\begin{equation}
\rho_{xy}(B)=R_0B +R_SM
\label{eq1}
\end{equation}
where the second term  $R_SM$ accounts for the AHE and the $R_S \propto \rho_{xx}$ or $\rho_{xx}^2$ for dominant skew scattering or side--jump scattering processes, respectively.

The proportionality $R_H -R_0 \propto \chi$ is expected in the incoherent high temperature regime with strongly anisotropic crystal field\cite{Kontani1997}. However, we have to mention that the data can be described similarly by $R_H = R_0 + C\rho\chi$, where $R_0$ is the normal Hall effect and $C$ a constant\cite{Fert1987}, as $\rho$ is weakly $T$--dependant in this temperature range. Thus, regardless of the underlying model, the Hall effect is dominated by the AHE in a large temperature range ($\unit[70]{K}-\unit[300]{K}$). $R_H$ is field independent (and so are $R_{0}$ and $R_{S}$), so it can be fitted using magnetic susceptibility data to obtain $R_0=\unit[-0.1]{\upmu\Omega\cdot cm/T}$ and $R_S=\unit[97]{\upmu\Omega\cdot cm/T}$ for the high temperature regime as shown in the inset in Fig.~\ref{Rh_noP}.

\begin{figure} 
	\centering
		\includegraphics{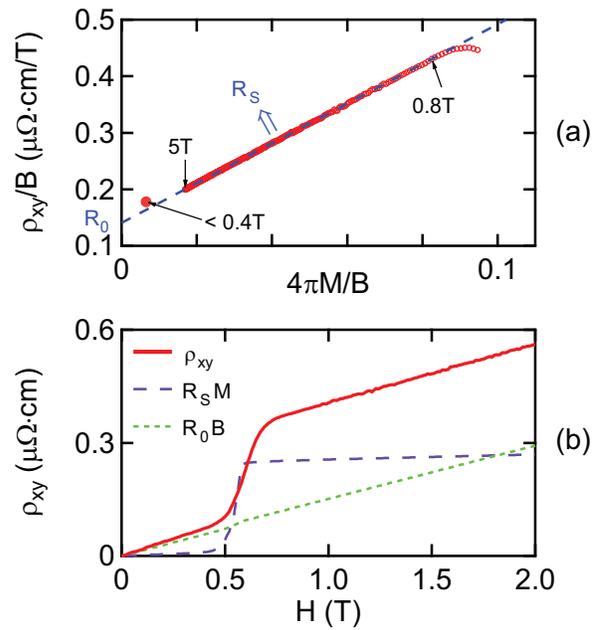}
		\caption{\label{rho_B_vs_M_B} \textit{(color online)} a) Determination of the coefficients $R_0$ and $R_S$ in the FM phase: $\rho_{xy}/B$ vs $4\pi M/B$ plot at $\unit[2]{K}$ ($P=0$) up to $\unit[5]{T}$, where $M$ is the magnetization density expressed in $\unit[10^{-4}]{emu\cdot cm^{-3}}$ and $B=H+4\pi M$ is the internal field in Tesla. The high--field part ($\unit[0.8]{T}<B<\unit[5]{T}$) is fitted  by eqn.~\ref{eq1} (dotted line) and we obtain $R_0=\unit[0.14]{\upmu\Omega\cdot cm/T}$ and $R_S=\unit[3.51]{\upmu\Omega\cdot cm/T}$. The low--field points ($0<B<\unit[0.4]{T}$) shrink altogether into a single point. The data $0.4<B<\unit[0.7]{T}$ is not shown here since it corresponds to a transition regime which has no physical interest (different samples for $\rho_{xy}$ and $M$ measurements). b) $\rho_{xy}$ and its constituent contributions $R_0B$ (NHE) and $R_SM$ (AHE) vs $H$ (external field), in the same experimental conditions and using the values of $R_0$ and $R_S$ obtained in (a).}
\end{figure}

The normal contribution to the Hall effect emerges only at low temperature. Figure \ref{rho_B_vs_M_B} gives an insight to the relation between Hall effect and magnetization at $\unit[2]{K}$, to separately account for the NHE and AHE contributions.

In the ferromagnetic region ($B>B_m$), at low temperature ($\unit[2]{K}$), both $\rho_{xy}(B)$ and $M(B)$ show only weak variations. Therefore, $R_0$ and $R_S$ are assumed to be field independent. Indeed, on a plot $\rho_{xy}(B)/{B}$ vs $M(B)/B$, see Fig.~\ref{rho_B_vs_M_B}(a), the high field data ($B>B_m$) fall on a straight line which can be fitted by eqn.~\ref{eq1} to obtain $R_0=\unit[0.14]{\upmu\Omega\cdot cm/T}$ and $R_S=\unit[3.51]{\upmu\Omega\cdot cm/T}$ in good agreement with Ref.~\citen{Matsuda2000}.
In the PM phase, since both $\rho_{xy}$ and $M$ increase linearly with field, the low--field points ($\unit[0]{T}<B<\unit[0.4]{T}$) shrink altogether into a single point. This point falls close to the FM line. The data $\unit[0.4]{T}<B<\unit[0.8]{T}$ corresponds to the metamagnetic transition regime. Obviously, there is no drastic change of $R_0$ between PM and FM phases through the metamagnetic transition, thus there is no direct evidence of a Lifshitz transition at $B_m$.

Figure \ref{rho_B_vs_M_B}(b) shows the field dependence of $\rho_{xy}$ at $T=\unit[2]{K}$. On the same scale, we plotted the contributions of $R_0B$ and $R_SM$ as determined in the FM phase shown in Fig.~\ref{rho_B_vs_M_B}(a), to compare the relative weight of NHE and AHE. The normal contribution $R_0B$ is basically linear with $H$ except for a small kink at $\unit[0.6]{T}$ due to the magnetization density which suddenly amounts to 10\% of the internal field ($4\pi M/B \approx 0.1$) when the system becomes ferromagnetic. But since the magnetization only increases very slowly in the FM phase, its consequence on the internal field is negligible, so that $B\approx \mu_0H$, the applied field. As a consequence, the slope of the $\rho_{xy}(B)$ curve is mainly determined by the NHE ($\partial\rho_{xy}(B)/\partial B\simeq  R_0$), and the step of $\rho_{xy}(B)$ at the transition is due to the AHE, mainly. In a simple approach, the jump in the AHE is due to a step--like increase of the magnetization ($\Delta\rho_{xy}\simeq R_S\Delta M$). However, from the experimental data we cannot exclude that also $R_S$ may be affected at the metamagnetic transition, as the resistivity $\rho_{xx}$ ---~at least in a longitudinal configuration~--- shows such a step--like increase. Finally, we should mention that the transition appears broader in $\rho_{xy}(B)$ than in $M(B)$. This difference marks the complex interplay between intrinsic and extrinsic (impurity) contributions to the AHE.


\section{Hall Effect under High Pressure}

\begin{figure}
	\centering
		\includegraphics{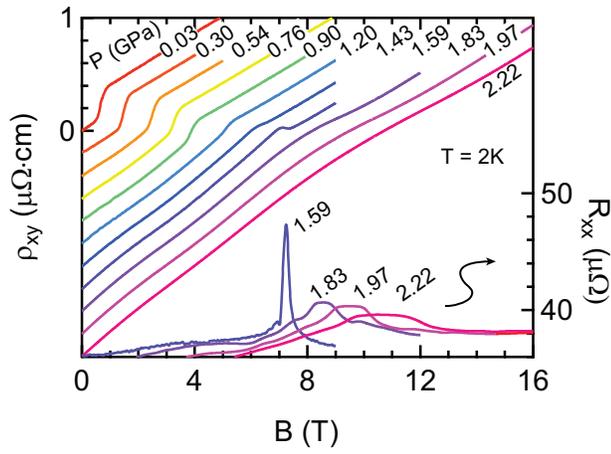}
	
	\caption{\label{hall_2K} \textit{(color online)} Pressure evolution of $\rho_{xy}$ vs $B$ at $\unit[2]{K}$ for different pressures (labels indicate the pressure in GPa). The curves are vertically spaced by $\unit[0.2]{\upmu\Omega\cdot cm}$ for clarity. In addition, we plot the symmetric part of the measured signal, the magnetoresistance $R_{xx}$, at $T=\unit[2]{K}$ as a function of $B$ for different pressures (1.59, 1.83, 1.97, and $\unit[2.2]{GPa}$, right scale).}
\end{figure}

\begin{figure}
	\centering
		\includegraphics{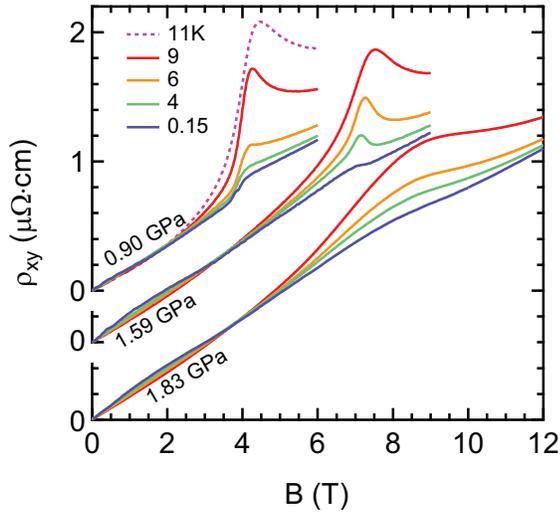}
	
	\caption{\label{P4P7P8} \textit{(color online)} $\rho_{xy}$  as a function of field at various temperatures ($0.15$ to $\unit[11]{K}$) under pressure ($0.90/1.59/\unit[1.83]{GPa}$). For clarity, curves at $1.59$ and $\unit[1.83]{GPa}$ are shifted vertically by $-0.4$ and $\unit[-1]{\upmu\Omega\cdot cm}$, respectively.}
\end{figure}

\begin{figure}
	\centering
		\includegraphics{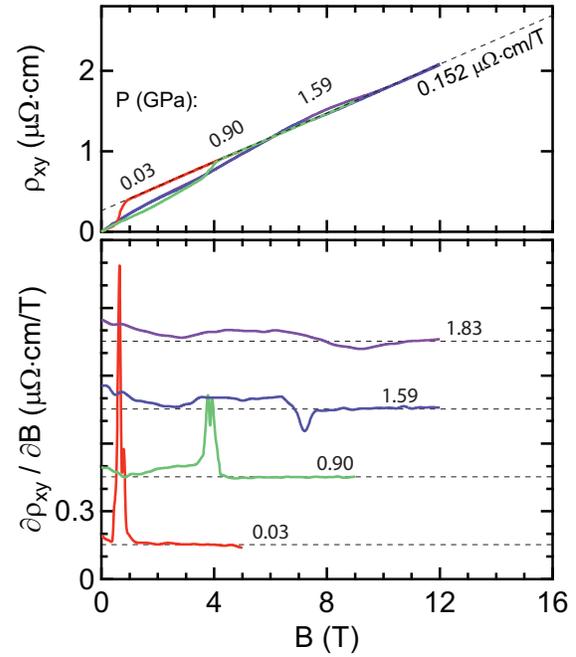}
		\caption{\label{hall_0K} \textit{(color online)} (upper panel) $\rho_{xy}$ vs $B$ for various pressures extrapolated to $T=0$  (labels indicate the pressure in GPa). The dotted line indicates the pressure independent linear increase of $\rho_{xy}$  in the ferromagnetic state above $B_m$ which has a linear slope of $\unit[0.152]{\upmu\Omega cm/T}$.   (lower panel) Field dependence of the differential Hall coefficient $\partial \rho_{xy}/\partial B$ for different pressures.} 
\end{figure}

In our pressure study, the Hall effect $\rho_{xy}(B)$ has been measured by performing field scans upwards at various temperatures in the range from $0.15$~K up to $\unit[20]{K}$. 
Figure \ref{hall_2K} represents $\rho_{xy}$ as a function of $B$ at $T=\unit[2]{K}$ for various pressures up to 2.22~GPa (the curves are vertically spaced by 0.2 $\upmu\Omega cm$ for clarity, respectively). While at low pressures ($P \leq \unit[1.2]{GPa}$) a clear jump of $\rho_{xy}$ at $B_m$ is observed, which is qualitalively similar to the behaviour at ambient pressure, above $\unit[1.6]{GPa}$ only a rather broad cross--over is observed. This is a first indication that the first order metamagnetic transition collapses in the pressure range from $\unit[1.2]{GPa}$ to $\unit[1.6]{GPa}$, thus the pressure of the quantum critical end point $P_{QCEP}$ is located in this pressure window. In agreement with previous measurements \cite{Aoki2011}, a strong field dependence of $\rho_{xx}$ is also detected at $T=\unit[2]{K}$. The magnetoresistance $R_{xx}$ shows a very sharp peak at $P=\unit[1.59]{GPa}$ while for higher pressures a plateau--like enhancement of $R_{xx}$ is observed.  On cooling, it was proved \cite{Aoki2011} and verified here that the simultaneous enhancement of the residual resistivity and the enhancement of the $A$ coefficient is correlated with crossing through the QCEP. 

Next we concentrate on the behaviour of $\rho_{xy} (B)$ at different temperatures which is shown in Fig.~\ref{P4P7P8} for $P=\unit[0.9]{GPa}$, $\unit[1.59]{GPa}$, and $\unit[1.83]{GPa}$. The Hall resistivity $\rho_{xy}$ at different temperatures at $\unit[0.9]{GPa}$ is rather similar to that at ambient pressure (see Fig.~\ref{Rh_noP}(b)) with a steplike anomaly at lowest temperatures and a peak which appears with increasing temperature just above $B_m$. At $P=\unit[1.59]{GPa}$ the behaviour appears likewise, except at lowest temperature (see below). For $P=\unit[1.83]{GPa}$ no sharp anomaly can be detected, but a broad crossover even at the lowest temperature. Furthermore, below $\unit[2]{K}$, the temperature dependence of the Hall resistivity is very small for $P<\unit[1.6]{GPa}$ which is close to $P_{QCEP}$. 

While low temperature data have been measured only in the pressure regime where the QCEP is expected, we extrapolated $\rho_{xy}(B)$ measured at finite temperature down to $T=0$ with a polynomial of second degree for all pressures. As we can see in Fig.~\ref{P4P7P8}, below $\unit[2]{K}$ the temperature dependence of the Hall resistivity is very small for $P<\unit[1.6]{GPa}$ ---~which is supposed to be close to $P_{QCEP}$~--- in such a way that the extrapolated $\rho_{xy}(B,T=0)$ almost coincides with the one measured at the lowest temperature.
Figure \ref{hall_0K} (upper panel) illustrates the pressure evolution of the Hall resistivity in the limit $T\to\unit[0]{K}$ for selected pressures in absolute values. Interestingly, in the entire pressure range, all the $\rho_{xy}(B)$ curves almost match together in the high field part ($B>B_m$). This clearly shows that the normal contribution to the Hall effect, which dominates the field dependence above $B_m$, as shown in Fig.~\ref{rho_B_vs_M_B}, is pressure independent in the polarized ferromagnetic regime and has a constant slope $\mathrm{d}\rho_{xy}/\mathrm{d}B = \unit[0.152]{\upmu\Omega cm/T}$ (indicated by the dashed line).  The lower panel of Fig.~\ref{hall_0K} shows the field dependence of the differential Hall coefficient $\tilde{R}_H=\partial \rho_{xy}/\partial B$ for selected pressures. At low pressure, a very sharp maximum marks the metamagnetic transition. With increasing pressure, the height of the maximum decreases, while the width stays constant. Apparently, $\mathrm{d}\rho_{xy}/\mathrm{d}B$ has changed its sign at $\unit[1.59]{GPa}$, while above $\unit[1.6]{GPa}$ only a broad crossover appears. The dashed lines in the lower panel of  Fig.~\ref{hall_0K} indicate a constant slope of $\mathrm{d}\rho_{xy}/\mathrm{d}B = \unit[0.152]{\upmu\Omega cm/T}$ which is pressure independent in the FM state above $B_m$.

In contrast to the FM phase, the Hall coefficient in the PM phase below $P_{QCEP}$ varies with field and pressure. Eventhough the $\rho_{xy}(B)$ curves for $P\leq \unit[0.30]{GPa}$ are linear in the low field PM phase, this is not the case at higher pressure, for $P\rightarrow P_{QCEP}$. 

\begin{figure}
	\centering
		\includegraphics{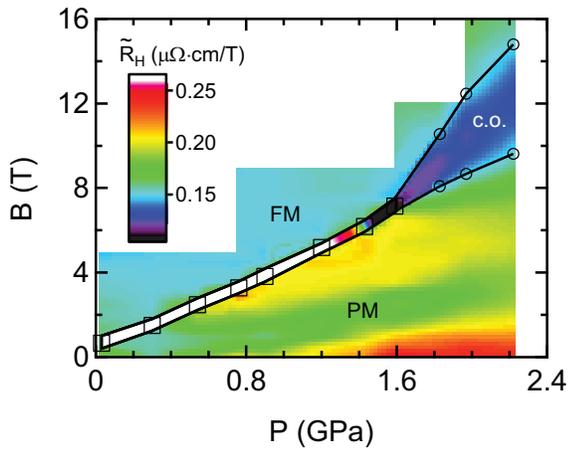}
		\caption{\label{P_H_colour} \textit{(color online)} Field--pressure phase diagram of UCoAl at $\unit[0]{K}$ obtained by Hall effect measurements. The colour scale maps the differential Hall coefficient $\tilde{R}_H=\partial\rho_{xy}/\partial B$ obtained from extrapolated curves of Figure \ref{hall_0K}. The bold lines delimit the width of the transition and the lower and upper limits of the crossover region (c.o.). The QCEP is located around $(\unit[1.6]{GPa},\unit[7]{T})$, above which the transition suddenly broadens.}
\end{figure}

To summarize the presentation of the experimental data, we plot in Fig.~\ref{P_H_colour} the differential Hall coefficient $\tilde{R}_{H}$ in a $(B,P)$ phase--diagram. It appears clearly that $\tilde{R}_H$ is never constant in the paramagnetic state while it varies little in the ferromagnetic state, as discussed above. 
The width of the step--like transition at $B_m$ is delimited by solid lines for $P<\unit[1.6]{GPa}$. One can reasonably affirm that the QCEP is located in the pressure window between the characteristic pressures $P_M = \unit[1.2]{GPa}$ and $P_\Delta = \unit[1.6]{GPa}$. At $P_M$, the differential Hall effect is changing sign, while above $P_\Delta$ the broadening of the crossover is clearly visible.  The robustness of the FM phase contrasts with the variations in the PM phase, in which a valley of low $\tilde{R}_H$ separates two regions of high $R_H$, one at low field and one at high field. It is interesting to observe that there is no drastic difference between the low pressure FM phase and the polarized paramagnetic phase (PPM) which is expected to occur for $P>P_{QCEP}$ through the crossover regime. As a matter of fact, the pressure decrease of the magnetization in the FM region has no effect in the high magnetic field regime.

The pressure dependence of $B_m$, defined by the extrema of $\tilde{R_H}$ shown in Fig.~\ref{P_H_colour} is in excellent agreement with previous transport measurements \cite{Aoki2011}. Magnetization measurements \cite{Mushnikov1999} led to lower pressure dependence of $B_m$, as mentioned above. This discrepancy probably comes from pressure inhomogeneities or differences in the crystal quality. The condition of good hydrostaticity may be crucial in UCoAl since the linear compressibility along the $a$--axis is known to be about $5$ times larger than that along the $c$--axis \cite{Havela2001,Maskova2012}.
In Figure \ref{hall_0K}, at low field, $\rho_{xy}$ appears $P$--independant. This observation is in good agreement with pressure magnetization data\cite{Mushnikov1999} and recent magnetostriction experiments\cite{Aoki2011} which led to the conclusion that the respective Gr\"uneisen parameter of the magnetization $\Omega_M=\frac{1}{M}\frac{\partial M}{\partial V}V$ is near $2$ for $B<B_m$ and $35$ for $B>B_m$, while $\Omega_{B_m}$, the Gr\"uneisen parameter of $B_m$, is near $140$.

\begin{figure}
	\centering
		\includegraphics{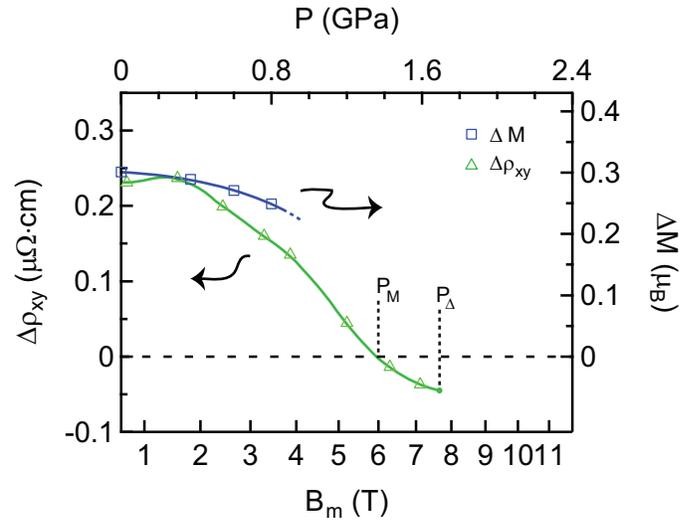}
		\caption{\label{deltarhoxy_deltaM} \textit{(color online)} Comparison of the jump in $\rho_{xy}(B)$ and $M(B)$ taken from Ref. \citen{Mushnikov1999}. Data are plotted versus $B_m$ to correct for discrepancy in pressure determination. Top axis indicates our corresponding pressure scale. Lines are guides for the eye. At $P_M$ the positive jump $\delta\rho_{xy}$ vanishes, and above $P_\Delta$ only a crossover from the PM to the field induced FM state occurs.} 
\end{figure}

\begin{figure}
	\centering
		\includegraphics{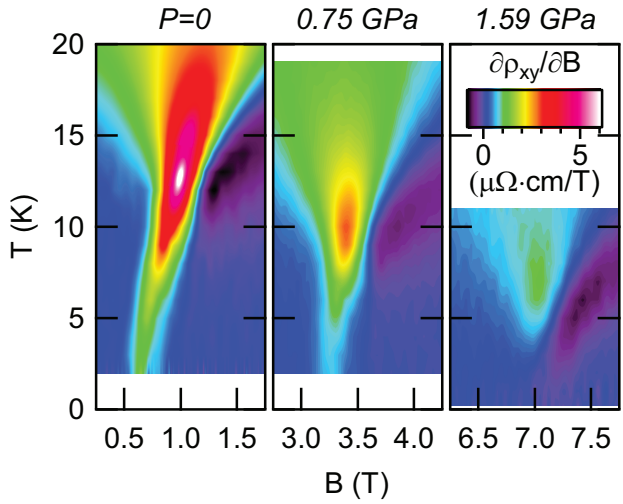}
	
	\caption{\label{colorphasediags} \textit{(color online)} $(B,T)$ phase diagram of UCoAl and its evolution under pressure as obtained by Hall effect measurements. The colour scale corresponds to $\partial \rho_{xy}/\partial B$. Each image was obtained from a mesh of field scans at $P=0$, $0.75$ and $\unit[1.59]{GPa}$.}
\end{figure}

\begin{figure}
	\centering
		\includegraphics{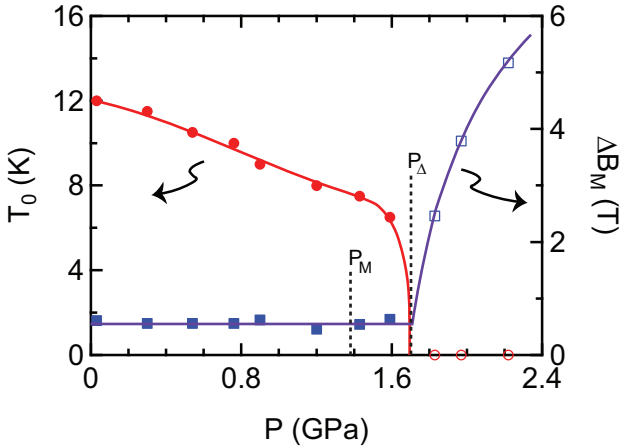}
\caption{\label{phase_diag} \textit{(color online)} Pressure evolution of the critical temperature $T_{0}$ (left axis) and the width of the transition at lowest $T$ (right axis), illustrating the clear termination of the transition at $P_\Delta \approx \unit[1.7]{GPa}$. Lines are guides for the eye.}
\end{figure}

Figure \ref{deltarhoxy_deltaM} shows the jump $\Delta \rho_{xy}$ at the metamagnetic transition plotted against the critical field $B_m$. It is compared to the corresponding evolution of the jump of the magnetization $\Delta M$ taken from Ref.~\citen{Mushnikov1999} which decreases slightly with pressure. In order to correct for differences either in $P$ determination, in hydrostaticity or in crystal quality, the data are plotted as a function of $B_m$.  (The upper scale indicates the correspondence with our $P$ determination).
 In a simple approach $\Delta \rho_{xy} \propto \Delta (R_0B) + \Delta (R_SM)$. At least up to $P_M = \unit[1.2]{GPa}$ the main contribution to $\Delta \rho_{xy}$ will be the jump in the magnetization $\Delta M$ at $B_m$. However, at $\unit[1.43]{GPa}$ and $\unit[1.59]{GPa}$, the corresponding anomaly in $\rho_{xy}$ at $B_m$ is no more a finite positive $\Delta \rho_{xy}$, but it appears slightly negative. Thus above $\unit[1.2]{GPa}$, the apparent signature of the first order metamagnetic transition is not directly detected in the Hall effect measurement. 
 Clearly, above $\unit[1.59]{GPa}$ no first order transition has been observed, indicating that the QCEP lies between $P_M=\unit[1.3]{GPa}$ and $P_\Delta = \unit[1.7]{GPa}$.  As already mentioned above, previous resistivity measurements pointed out the strong increase of the inelastic term, namely the $A$ coefficient. Obviously, the associated change of the effective mass has a strong influence on the Hall resistivity. 

Finally, in Fig.~\ref{colorphasediags} we show $\partial\rho_{xy}(B,T)/\partial B$ for different pressures as a function of magnetic field and temperature for different pressures in a colour plot. The evolution of  $\tilde{R}_H (T,B)$  with increasing pressure can be described in simple terms. At ambient pressure, the position of the critical end point is well defined by a 
sharp maximum in $\tilde{R}_H$. With increasing pressure, the amplitude of $\tilde{R}_H$ at the critical end point decreases, $T_0$ decreases, and the critical field increases. Figure \ref{phase_diag} shows the pressure evolution of the temperature of the critical end point $T_0$ (defined by the maximum of $\tilde{R}_H$) and the width of the metamagnetic transition $\Delta B_m$ (width of the peak in $\partial \rho_{xy}/\partial B$). Surprisingly, $T_0$ is not suppressed continuously to zero temperature but rather drops from $T_0 = \unit[6]{K}$ to zero in the vicinity of $P_\Delta=\unit[1.7]{GPa}$. This pressure also corresponds precisely to the end of the sharp metamagnetism, unambiguously attested by the sudden increase of the transition width $\Delta B_m$ on entering a crossover region beyond the QCEP.

\section{Discussion}

The Hall effect experiments give detailed insights on the behaviour of the metamagnetic transition in the itinerant heavy fermion system UCoAl. 
For $T\ll T_0$, the step--like feature of the $\rho_{xy}(B)$ curves is connected to the magnetization jump via the anomalous Hall effect. The constant Hall coefficient in the FM phase at low temperature (field-- and pressure--independant) strongly suggests that the coefficients $R_0$ and $R_S$ do not change with pressure and are the same as at ambient pressure, where they are known with precision. On approaching the QCEP, a drastic change occurs in the Hall signal on crossing $B_m$. We associate this to the enhancement of the average effective mass detected via the $A$ coefficient in resistivity measurements. 

For Ising--type Uranium ferromagnets, as the jump of the magnetization at $B_m$ is often strong (near $\unit[0.5]{\mu_B}$) and the renormalized Fermi energy is low, Fermi surface reconstruction can be expected at $P_c$ and $B_m$ as clearly observed in UGe$_2$ \cite{Settai2002}. In the case of UCoAl, at least at low field ($B \ll B_m$) and at high field ($B > B_M$), no major variation of the Hall constant can be pointed out. There is no evident signature of a Fermi surface reconstruction on sweeping from the PM to FM phase for $P < P_{QCEP}$ and from the PM to a polarized paramagnetic phase for  $P > P_{QCEP}$. Recent thermoelectric power experiments at ambient pressure were interpreted only via the decrease of the heaviest hole effective mass through $B_M$ \cite{Palacio-Morales2013}. There was also no signature of a Lifshitz transition. By contrast, in the case of the ferromagnetic superconductor UGe$_2$, the crossing from PM to FM is accompanied by a sign change of the Hall coefficient in excellent agreement with a Fermi surface reconstruction detected in quantum oscillations experiments \cite{Haga2002}. However, in these multiband heavy fermion systems, the Hall resistivity response can be complex. A pathological example is the pseudo--metamagnetic transition in CeRu$_2$Si$_2$ where no clear signature of a Fermi surface reconstruction is detected in the Hall constant \cite{Kambe1996} while there is direct evidence of a Fermi surface change through $B_m$ by quantum oscillations experiments\cite{HAoki1993} associated with a Lifshitz transition. So the question remains open for UCoAl.

It is interesting to recall the case of the ferromagnetic itinerant system $\mathrm{ZrZn_{2}}$ ($T_{Curie}\sim \unit[28.5]{K}$, $M_0 \sim \unit[0.17]{\mu_B}$ per Zr atom), which is an isotropic Heisenberg system. As in $\mathrm{UGe_{2}}$, evidences are found for two emerging FM phases (FM$_2$,FM$_1$) at low pressure ($P_C \sim \unit[1.65]{GPa}$) and one field--induced FM$_1$ phase when a field is applied from the PM phase above $P_c$. However, at least for FM$_1$, it seems that $P_{QCEP}$ almost coincides with $P_c$ \cite{Uhlarz2004}. As a consequence, at $\unit[2.1]{GPa}$, the crossover metamagnetic field is so low ($\unit[0.05]{T}$) that the observation of the Fermi surface in the PM regime is proscribed \cite{Kimura2004}. De Haas -- van Alphen measurements pointed out the crossover between the FM$_1$ and FM$_2$ phases with an invariance of the Fermi surface topology. The change in the exchange splitting of the Fermi surface is the signature of the field crossover FM$_1$/FM$_2$. Let us point out that in the case of $\mathrm{ZrZn_{2}}$ the Fermi energy is weakly renormalized, thus the local fluctuations are negligible compared to the case of U compounds where the itinerant--localized duality is a major part of the puzzle.

Finally, a striking point in our results is the appearance of different field regimes in the low field PM response ($B<B_m$) on approaching $P_{QCEP}$. It may be connected to the particularity of the quasi--Kagome structure of UCoAl, leading to a complex interplay between frustrated magnetic fluctuations and sole FM interactions. Other evidences of non--conventional FM character of UCoAl is the observation of non--Fermi liquid properties in the PM regime at pressure far below $P_{QCEP}$ \cite{Palacio-Morales2013}. To solve the UCoAl puzzle, key experiments are now to extend the magnetization measurements through the QCEP and of course to succeed to directly observe the Fermi surface.

\section{Conclusions}
Hall effect experiments at ambient and at high pressure have been presented in the itinerant metamagnet UCoAl. At the metamagnetic transition $B_m$, the jump of the Hall effect is dominated by the AHE contribution which scales with the magnetization. The jump at $B_m$ disappears at $P_M \approx \unit[1.3]{GPa}$, and above $P_\Delta \approx \unit[1.7]{GPa}$ only a broad pseudo--metamagnetic transition could be detected. While the Hall effect in the polarized ferromagnetic state seems to be pressure independent, the interpretation of the field dependence in the paramagnetic state below $B_m$ under pressure on approaching the QCEP appears less straightforward and needs to be revised in the future. Finally, our Hall effect experiments do not allow to conclude on a possible Fermi surface change through the QCEP.

\section{Acknowledgements}

We thank H.~Harima, P.~Haen, H.~Kotegawa, K.~Ishida, and S.~Araki for useful discussions. This work was supported by ERC starting grant (NewHeavyFermion), and French ANR project (SINUS).



\end{document}